\documentclass[prc,aps,onecolumn,showpacs]{revtex4}
\usepackage{graphicx}
\newcommand{\be}{\begin{equation}}
\newcommand{\ee}{\end{equation}}

\newcommand{\ovl}{\overline}

\begin{document}
\title{Search for Higher Flavor Multiplets
in Partial Wave Analyses}

\author{Ya.~I.~Azimov$^{a,b}$, R.~A.~Arndt$^c$, I.~I.~Strakovsky$^c$,
R.~L.~Workman$^c$, K.~Goeke$^d$}

\affiliation{$^a$Petersburg Nuclear Physics Institute,\\
                 St.~Petersburg, 188300 Russia\\
             $^b$Thomas Jefferson National Accelerator Facility,\\
                 Newport News, VA 23606, USA\\
             $^c$Center for Nuclear Studies, Physics Department,\\
                 The George Washington University, \\
                 Washington, DC 20052, USA\\
             $^d$Institute for Theoretical Physics II,\\
                 Ruhr University, Bochum, 44801, Germany}

\bigskip
\begin{abstract}

The possible existence of higher multi-quark flavor multiplets 
of baryons is investigated.  We argue that the $S$-matrix should
have poles with any quantum numbers, including those which are
exotic.  This argument provides a novel justification for the
existence of hadrons with arbitrary exotic structure.  Though it
does not constitute a proof, there are still no theoretical
arguments against exotics.  We then consider $KN$ and $\pi N$
scattering.  Conventional and modified partial-wave analyses
provide several sets of candidates for correlated pairs
$(\Theta_1,\,\Delta)$, each of which could label a related
\textbf{27}-plet. Properties of the pairs (masses, mass
orderings, spin-parity quantum numbers) do not quite correspond
to the current theoretical expectations. Decay widths of the
candidates are either wider or narrower than expected. Possible
reasons for such disagreements are briefly discussed.
\end{abstract}


\maketitle

\section{Introduction}

Until very recently, it was commonly believed that all hadrons had
only ``minimal'' quark structure. This meant that every meson could
be considered as composed of only one quark-antiquark pair (or,
perhaps, of gluons, without quarks at all), with every baryon
composed of three quarks.  This hypothesis implied a restriction on
the possible quantum numbers of the hadrons (flavor numbers, in
particular).  However, nobody could suggest, in QCD or through other
approaches, a mechanism to forbid hadrons containing additional
quark-antiquark pairs and having exotic quantum numbers, incompatible
with the ``minimal" structure. This point was especially hot, since
due to virtual gluon radiation and quark-antiquark pair production
(\textit{e.g.}, by vacuum polarization) every hadron has some
probability to be seen in a configuration with (very) many quarks 
and gluons. Such configurations should emerge as short-time 
fluctuations of the initial state, even if it is ``minimal". 
Therefore, they may play the leading role in the hard (\textit{i.e.}, 
short-time) processes. Indeed, such hard processes as deep-inelastic 
lepton scattering, Drell-Yan pair production, and so on, are known 
to be well described by structure functions, based on the presence 
in a hadron of both the constituent quarks (corresponding to the 
``minimal" set) and the infinite number of ``sea" quark-antiquark 
pairs.

The first attempts to theoretically understand the properties of
multi-quark states (mainly, of mesons) were made in the framework
of the MIT bag model~\cite{jaf}. Later, another method became
popular for baryons. This approach is related to the chiral soliton
model ($\chi$SM; see recent reviews~\cite{wk,elkp} for more 
detailed description and references), and has allowed the clear
prediction~\cite{dpp} of a new baryon state with unique properties.
New experimental searches, stimulated by the prediction, have
provided evidence for the first exotic baryon, the $\Theta^+$ with
mass $M_{\Theta^+}$ about 1540~MeV and strangeness $S=+1$.  The
earliest positive data were obtained by the collaborations
LEPS~\cite{leps}, DIANA~\cite{diana}, and CLAS~\cite{clas}. Now,
more than 10 publications support existence of the $\Theta^+$, 
with decays to both $K^+n$ and $K_Sp$. Additional evidence for 
the $\Theta^+$ (or some other exotic baryon(s) with $S=+1$) has 
been demonstrated recently~\cite{KA} in properties of $K^+$-nuclear
interactions.

Not all searches have yielded positive results.  Some collaborations
have not (yet) found the $\Theta^+$ in their data. Of these negative
results, some have been formally published (see, \textit{e.g.},
Refs.~\cite{BES,HB,Phen,Lands,babar,cdf,hCP,LEP,belle,aleph,FOCUS}),
while others exist mainly as rumors, or as conference slides.
Nevertheless, all of these cast doubt on the existence of the
$\Theta^+$. Note that the negative results mainly correspond to
higher energies than positive ones, and could be determined by
different mechanisms. A new set of dedicated experiments,
performed by several independent groups, are rather soon expected 
to provide more clear conclusions on the existence of this and 
other exotic hadrons.

The most decisive experimental check of the $\Theta^+(1540)$ will 
come from good measurements of $K^+n$ and/or $K_Lp$ elastic
scattering and/or charge exchange. Present data on these processes
have insufficient quality and are only able to provide an upper
bound for $\Gamma_{\Theta^+}$ (not more than
$\sim1~{\rm MeV}$)~\cite{nus,asw,ct,aswm}. In new precise
measurements, with good resolution, the $\Theta^+(1540)$ should
manifest itself as a narrow peak, for which the height (up to the
experimental resolution) may be calculated in an absolutely
model-independent way. Such data would therefore allow one to
either definitely confirm, or definitely exclude the
$\Theta^+(1540)$.

More detailed analysis of the existing data shows that, though the
present non-observation data require exotic production to be small
as compared to conventional hadrons, they do not entirely exclude
the existence of the $\Theta^+$ and/or its companions/analogs. For
example, analysis of the BES data~\cite{BES}, presented in
Ref.~\cite{as}, demonstrates some suppression of the
$\Theta$-production. However, given the present experimental
accuracy, this suppression is not severe, an essentially stronger
suppression of the exotic production could still have a natural
explanation.  Similar conclusions apply also for other data sets 
(see, \textit{e.g.}, Ref.~\cite{babar}). For this reason, we will 
assume the $\Theta^+$ (as well as other multi-quark hadrons) to 
exist, and will discuss the consequences.

If all baryons were, indeed, classified as three-quark systems,
then only a very limited set of flavor multiplets would be possible,
corresponding to the product of three triplet representations
\textbf{3} of the flavor group $SU(3)_F$. Since
\be
\label{3t3}
{\bf3}\times{\bf3}\times{\bf3}={\bf1}+2\cdot{\bf8}+{\bf10}\,,
\ee
any baryon, in such a case, should either correspond to a flavor
singlet \textbf{1} (this is possible only for $\Lambda$-like 
baryons), or be a member of an octet \textbf{8} or a decuplet 
\textbf{10}.

A special role in this picture is played by non-strange baryons.
Nucleon-like baryon can not form a unitary singlet, nor can it be
member of a decuplet. Therefore, in the conventional three-quark
picture of baryons, nucleon-like baryon can only be the member of
an octet. It is natural, further, to suggest that every existing
$SU(3)_F$-multiplet contains all possible states, even though the
symmetry is violated and the mass degeneracy is broken. Then 
each $N$-like state should be accompanied by the whole set of 
associated octet states.

Thus, if only three-quark baryons existed, any $N$-like baryon
could be used as a label for the corresponding flavor octet of
baryons. In the same way, any $\Delta$-like baryon would be
unambiguously related to the accompanying decuplet of baryons,
and, therefore, could be a label for the corresponding set of
states. In this short discussion, we have not accounted for a
possible mixing between states of different multiplets. However, 
it is not essential for our present purpose of state counting: 
the mixing changes relations between different hadrons (relations 
of masses, of effective coupling constants and so on), but does 
not change the number of states.

\begin{figure*}[th]
\centering{
\includegraphics[height=0.7\textwidth, angle=90]{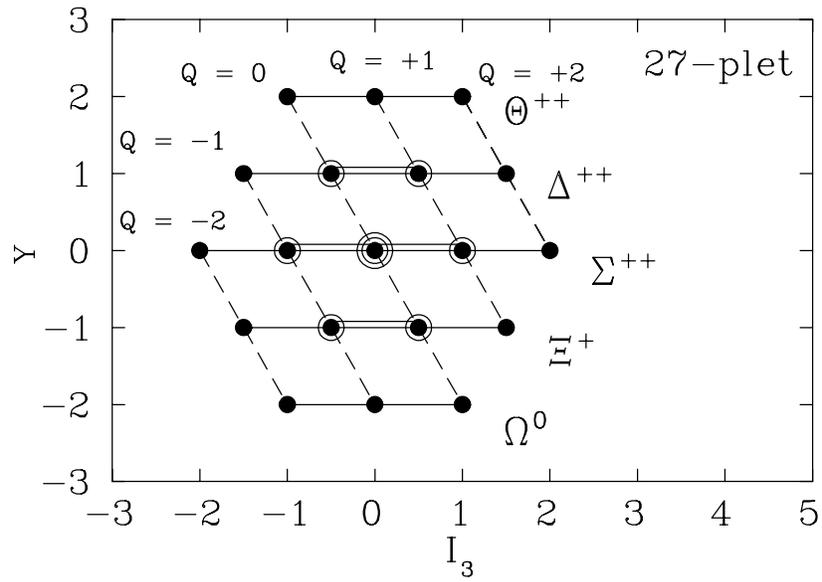}
}\caption{Structure of a possible baryon \protect\textbf{27}-plet,
          for non-violated $SU(3)_F$. Hypercharge $Y$ for baryons
          is $S+1$. Full circles correspond to non-degenerate states.
          Each additional open circle corresponds to an additional
          state with the same values of $Y$ and $I_3$, but with
          different value of $I$. Isotopic multiplets (constant
          values of the charge) are shown by solid (dashed) lines.
          Explicitly written for each value of $Y$ are the baryons
          with the largest value of charge.}
          \label{fig:g1}
\end{figure*}

If exotic hadrons do really exist, the situation becomes different
from one considered above. Five-quark states may provide new flavor
multiplets, according to the product relation
\be
\label{5t3}
{\bf3}\times{\bf3}\times{\bf3}\times{\bf3}\times{\bf\ovl3}
=3\cdot{\bf1}+8\cdot{\bf8}+4\cdot{\bf10}+2\cdot{\bf\ovl{10}}+
3\cdot{\bf27}+{\bf35}\,,
\ee
If, for instance, the antidecuplet(s) $\ovl{{\bf10}}$ do(es) exist,
some nucleon-like states may belong to antidecuplet(s), instead of
octets. If even higher multiplets \textbf{27} and \textbf{35} exist,
they should contain $\Delta$-like states not related with any
decuplet. Moreover, the \textbf{27}-plet, as well as many other (but
not all) higher multiplets, contains both $N$-like and $\Delta$-like
members (its multiplet diagram is shown in Fig.~\ref{fig:g1}; see
also Ref.~\cite{deSw} for some higher multiplet diagrams). The
\textbf{35}-plet, which may also be composed of 5 quarks, would have
no $N$-like member (similar, in this respect, to the decuplet
\textbf{10}). Instead, it would contain, in addition to the
$\Delta$-like state, a new kind of non-strange states, having 
isospin $5/2$. Up to now, such states have not been explored at all.

Thus, existence of $\ovl{{\bf10}}$ and/or higher multiplets
prevents $N$'s and $\Delta$'s from being good labels for flavor
unitary multiplets. Instead, the higher multiplets have another
characteristic member, a baryon with positive strangeness - the
$\Theta^+$.

There is one crucial difference between $\ovl{{\bf10}}$ and even
higher multiplets. Being a member of $\ovl{{\bf10}}$, $\Theta^+$ is
an isosinglet and has no positive-strangeness partners. In \textbf{27},
\textbf{35}, and other higher flavor multiplets, the $\Theta^+$ should
be a member of a non-trivial isospin multiplet. Hence, it should be
accompanied by isospin partners, also having $S=+1$. Each baryonic
\textbf{27}-plet contains isotriplet $\Theta_1$, with the
subscript denoting its isospin $I=1$.  It consists of $\Theta^0,\,
\Theta^+$, and $\Theta^{++}$. Each \textbf{35}-plet contains an
isomultiplet $\Theta_2$, with $I=2$, consisting of 5 members. In
both cases, one of the isomultiplet members should be the
double-charged exotic baryon $\Theta^{++}$. Note, that no
experiment gives evidence for the $\Theta^{++}$ with mass near the
$\Theta^+(1540)$, thus supporting its isospin $I=0$ and its
expected antidecuplet nature. But if any $\Theta^{++}$ exists, as a
member of \textbf{27}, \textbf{35}, or even higher flavor
multiplet, it should, in its turn, have the double-charged
non-strange partner(s), $\Delta^{++}$ with $I=3/2$ (for \textbf{27},
see Fig.~\ref{fig:g1}), or even two such states, corresponding to
$I=3/2$ and $I=5/2$ (for \textbf{35}).

From this brief discussion, we see that the $\Theta^{++}$ could be
used as a evidence for multiplets higher than $\ovl{{\bf10}}$.
Note, however, that
\be
\label{2t8}
{\bf8}\times{\bf8}={\bf1}+2\cdot{\bf8}+{\bf10}
+{\ovl{\bf10}}+{\bf27}\,.
\ee
Since the ``stable'' (\textit{i.e.}, having no strong decays) mesons
and baryons (except the heavy $\Omega^-$-baryon) belong to flavor 
octets, we see, that among multi-quark multiplets only members of 
${\ovl{\bf10}}$ and/or \textbf{27} can decay to a pair of such 
``stable'' hadrons or appear as a resonance in their scattering. For 
instance, the isosinglet $\Theta^+$, a member of ${\ovl{\bf10}}$, 
can decay to $KN$. The $\Theta^{++}$, as seen in $K^+p$ interaction, 
evidently can not have $I=0$, as should be for the $S=+1$ member of 
${\ovl{\bf10}}$.  But if isospin violation is negligible, it can 
not also have $I=2$, as would be necessary for a \textbf{35}-plet. 
This implies the impossibility to search for \textbf{35} or further 
higher flavor multiplets in 2-hadron processes (in particular, in 
meson-baryon elastic scattering). The same is true for processes of 
electro- or photoexcitation of the nucleon, where the photon appears 
as the member of an $SU(3)_F$ octet, and for 2-hadron mass 
distributions in multi-hadron final states.

That is why, in what follows, we will discuss mainly \textbf{27}-plet,
considering it as an example of higher flavor multiplets. A
characteristic feature of the \textbf{27}-plet is the presence of 
both $\Theta^{++}$ and $\Delta$ members. They must be correlated, in
the sense that their spin and parity should be the same; their
masses may differ, since the $SU(3)_F$ is violated, but we do not 
expect them to be widely separated.

Of course, the \textbf{27}-plet should also contain an $N$-like 
member.  However, Eq.~(\ref{5t3}) shows that the 5-quark system may 
reveal numerous $N$-like states, not related to \textbf{27}-plets 
(one for each \textbf{8} and/or ${\ovl{\bf10}}$). Therefore, the 
pair correlation of $\Theta^{++}$ with $N$ (or triple correlation 
of $\Theta^{++},\,\Delta,\,N$) is less characteristic than the pair
correlation of $\Theta^{++}$ with $\Delta$. So, generally, we will
not look for it here.

In the next Section, we give a new argument for existence of
$S$-matrix poles with any quantum numbers, including exotic ones.
We also discuss which poles can be related to physical resonances.

In the Section~III, we examine existing data from $KN$ and $\pi N$
scattering, with $I=1$ and 3/2 respectively, to search for possible
$\Theta^{++}$ and $\Delta$ candidates with the same values of $J^P$
and with correlated masses. We will investigate the possibilities 
of both wide and narrow states.

The obtained results and their meaning are discussed in the
concluding Section~IV.

\section{Resonance spectroscopy and complex-energy poles}

As is well-known, stable particles are related to poles in energy
for some scattering amplitudes. These poles appear at real
energies below the lowest physical threshold of the corresponding
scattering channel (for the relativistic description, it is
more convenient to use not the energy itself, but the c.m. energy
squared). By analogy, we will consider resonances (unstable bound
states) as poles at complex values of the energy. In what follows,
we will study resonances, in particular, in physical channels with
exotic quantum numbers. Up to now, the mere existence of such 
states has been doubted. As a result, we begin with discussion of 
this issue.

Here we encounter two sets of questions. Some of these may be
called theoretical: How many (if any) complex poles may an
amplitude have?  Which poles may be considered as corresponding to
resonances?  Other questions are more phenomenological. They are
related to the search for amplitude poles starting from 
experimental data. These sets of questions are physically similar, 
though they look and are treated differently. Let us consider them 
in more detail.

\subsection{Theoretical questions}

It is the standard assumption that the strong interaction amplitudes
have definite analytical properties. For the amplitudes of
2-particle-to-2-particle processes they provide dispersion relations
in both the energy and momentum transfers. Most such relations
have never been formally proved on the basis of general axioms from
Quantum Field Theory (a rare exception is the energy dispersion
relation for pion-nucleon forward scattering). Quantum Chromodynamics, 
which is believed to underlie strong interactions, also cannot be 
used today to prove the dispersion relations for hadronic amplitudes, 
since even the transition from quarks and gluons to hadrons has not 
yet been traced without any additional assumptions and in a 
model-independent way. Nevertheless, such relations are widely used 
in the phenomenological treatment of strong interactions. For 
instance, analytical properties are assumed to extract meson-meson 
elastic amplitudes from data on processes involving meson-nucleon 
transformation into 2-meson-nucleon. Various dispersion relations 
are used now as input in modern Partial-Wave Analyses (PWA; see, 
\textit{e.g.}, the latest $\pi N$ PWA~\cite{piN}). They are also
applied to extract such strong interaction parameters, as 
meson-nucleon coupling constant(s) and the so called $\sigma$-term. 
Up to now, the application of dispersion relations has not 
induced any inconsistencies. Therefore, it seems reasonable to 
investigate consequences of the dispersion relations in more detail, 
in particular, with respect to the problem of exotic states.

The assumption of the momentum-transfer dispersion relations for
strong interaction amplitudes of 2-particle-to-2-particle
hadronic processes implies the possibility of analytical
continuation for the corresponding partial-wave amplitudes
to complex values of angular momentum $j$ (the Gribov-Froissart
formula~\cite{GF}, see also the monograph~\cite{Col}).

A strong interaction amplitude may have poles in the complex energy
plane, each with definite $j$ (which physically takes only discrete
integer or half-integer values). Instead of such energy-plane poles,
one can consider Regge poles, in the complex $j$-plane, their positions
and residues being dependent on energy. Positions of the two kinds of
poles are connected by a relation of the form $$F(E,\,j)=0\,,$$ and
have one-to-one correspondence. Therefore, we may apply the above
questions to the Regge poles, instead of the energy-plane poles. Such
an approach allows us to answer the first question, about the number of
poles.

As was shown by Gribov and Pomeranchuk~\cite{GP}, when the energy
$E$ tends to a value of $E_{\rm th}$, the (elastic or inelastic)
threshold of two spinless particles, the Regge poles accumulate to
the point
\be
\label{l12}
j=l=-1/2\,.
\ee
Here $j$ is the total angular momentum, while $l$ is the orbital
angular momentum of the two interacting particles (of course, they
generally differ, but coincide for the system of two spinless
particles). The movement of accumulating Regge poles near the
threshold is described by the trajectories
\be \label{grpom} l(E)\approx -\frac12
+ \frac{i\pi n}{\ln(R\sqrt{-k^2})} +{\cal O}(\ln^{-2}(R\sqrt{-k^2}))
\,, \ee
with $R$ and $k$ being the effective interaction radius and
relative c.m.
momentum. The number $n$ takes any positive and negative integer
values, $n=\pm1,\pm2,...\,$. Just below the threshold (at $k^2<0$)
the first two terms of Eq.~(\ref{grpom}) describe infinitely many
pairs of reggeons, which are complex conjugate to each other and tend
to the accumulation point at $k^2\to 0$. Note that relativistic
amplitudes always have many thresholds, for production of two- or
many-particle intermediate states. More detailed investigation of
the correction
terms in Eq.~(\ref{grpom}) shows that the simple pair-wise complex
conjugation of the poles is exactly true near (below) the lowest
threshold,
but is approximate near the higher thresholds, being violated by
small correction terms. This complex conjugation ordering becomes
destroyed also just above the threshold (even the lowest one), where
$k^2>0$ and $\ln(-k^2)$ is complex.

These results are quantitatively not general: for the threshold of
two particles with spins $\sigma_1$ and $\sigma_2$ there appear
several accumulation points~\cite{az}, the rightmost one is at
\be \label{j12}
j=-1/2+\sigma_1+ \sigma_2\,,
\ee
instead of (\ref{l12}). The reason is simple: the accumulation
points still correspond to $l=-1/2$, but particle spins provide several
possible $j$-values for any fixed $l$-value, and \textit{vice versa}.
Correspondingly, structure of the accumulations is still described by
trajectories (\ref{grpom}), with the shifted limiting points in the
$j$-plane. The case of multi-particle thresholds has never been really
explored (though hypothesized by Gribov and Pomeranchuk~\cite{GP}).

The qualitative nature of the results is general enough.  The
accumulations are directly related to the threshold behavior
\mbox{$\sim (k_i)^{l_i}(k_f)^{l_f}$}, characteristic for the strong
interaction amplitudes without massless exchange contributions
(here $k_i$ and $k_f$ are the relative c.m. momenta, $l_i$ and
$l_f$ are the orbital momenta, and the subscripts $i$ and $f$
correspond to the initial and final states).

The Gribov-Pomeranchuk (GP) accumulation phenomenon is not a
specific property of relativistic amplitudes; it emerges also
in the case of non-relativistic potential scattering with the finite
range $R$ of interaction, where the scattering amplitude has the
threshold behavior \mbox{$\sim (kR)^{2l}$} (the accumulation was
explicitly demonstrated~\cite{aas1}, in particular, for the Yukawa
potential $V(r)= g\exp(-\mu r)/r$, having the effective interaction
radius of order $1/\mu$).

Essential for us now is the infinite number of the accumulating
Regge poles~\cite{GP}, which implies that the total number of the
Regge poles is certainly infinite as well. For the partial-wave
amplitude of two-particle interaction with the fixed angular
momentum, this corresponds to the infinite number of poles in the
energy plane (all Riemann sheets). For the non-relativistic case of
the Yukawa potential, in the limit $\mu\to0$ (transforming the
Yukawa potential into the Coulomb one, with an infinite radius of
interaction), the infinite number of energy-plane poles is seen as
the infinite number of Coulomb radial excitations. In such a limit,
the GP accumulation of Regge poles near the threshold takes the form
of accumulation of Coulomb bound states to the threshold~\cite{aas1}
(note that the double limiting transition $\mu\to0,\, k\to0$ is not
equivalent here to the similar, but reversed limit $k\to0,\,\mu\to0$).

We emphasize that the above arguments have not assumed specific
quantum numbers in the scattering channel. Therefore, their
consequences should be equally applicable (or non-applicable) to
both bosonic and fermionic hadron poles, with any flavor quantum
numbers (exotic or non-exotic).

Thus, if we study $2\to2$ strong interaction amplitudes, we should
admit the existence of (an infinite number of) complex-energy poles
with any exotic quantum numbers, both mesonic and baryonic.
Alternatively, one could assume that analytical properties of
amplitudes having exotic quantum numbers for at least one of physical
channels ($s$-, $t$-, or $u$-channels) are essentially different from
those of totally non-exotic amplitudes.  However, we consider the
latter case to be unnatural. Thus, we obtain one more argument for
existence of exotics, absent in the previous publications.

Several earlier arguments for exotics were collected and briefly
discussed in Ref.~\cite{as}. One of them also uses reggeons and, thus,
might look similar to the present result. However, they are
essentially different. That older evidence for exotics~\cite{Ros}
was based on ``duality" of resonances and reggeons, which was
understood as equivalence between the sum of resonances in the
direct channel ($s$-channel) and the sum of reggeons in the
exchange channel ($t$- or $u$-channel), see Ref.~\cite{Col}. The
arising evidence for exotics can be avoided, \textit{e.g.}, by
assuming a ``conspiracy" of resonances or reggeons.  Our reasoning
is different, using reggeons in the direct channel and their
one-to-one correspondence with familiar energy poles in the same
channel. The result in this case cannot be so easily circumvented.

Note, that all suggested arguments for exotics, either theoretical
or phenomenological, give evidence, but cannot prove existence.
We emphasize here, however, that no firm theoretical arguments
have been presented to forbid exotic hadrons. Until either exotics
are observed, or their absence is understood, one cannot be
convinced that the present picture of strong interactions is
complete and self-consistent.

Now we encounter another theoretical question: whether every
complex-energy pole should be considered as possibly related with a
resonance? The problem is that, in terms of complex angular momenta,
the Regge poles participating in the GP accumulations are clearly
separated from the physical points, which are, for the $j$-plane,
only the integer (or half-integer) non-negative points.  Thus, these
poles cannot provide physically meaningful (bound or resonance)
states, at least near the corresponding threshold.

It might be that the total set of Regge poles is split into two (or
more?) different subsets: one related, say, with the GP accumulations,
another with the bound and/or resonance states. Then there could be
an infinite number of poles of the former type, while only few (if
any) poles of the latter type. If so, there could appear no bound or
resonance states with exotic quantum numbers, though there are
infinitely many poles somewhere on Riemann sheets of the energy plane.

Such a problem was also investigated for the Schr\"odinger equation
with the Yukawa potential~\cite{aas2}. Results obtained there show
no basic difference between various Regge poles: all Regge
trajectories appear to be different branches of the same
multi-valued analytical function. Formally, this fact is related
to non-trivial analytical properties of the Regge trajectories
as functions of the energy: their singularities come not only
from physical thresholds, but also from coincidence of two (or
more) reggeons~\cite{aas2}, where those reggeons can be interchanged.

The non-relativistic scattering off the Yukawa potential seems to
provide a good test-ground for analytical properties of the $2\to2$
relativistic scattering amplitudes. Thus, it is natural to think
that all poles in the energy plane at fixed angular momentum
for relativistic amplitudes also have a common nature. Then the
notion of a resonance becomes rather conventional: \textit{the pole
to be considered as related to a resonance state should be placed
not too far from the physical region}, to produce noticeable
enhancement in the physical amplitudes. This means, in particular,
that the resonance width should not be too large. Quite conventionally,
today we may take, say, $\Gamma^{tot}<500$~MeV. Such a boundary may be
increased in future, when both experimental precision and theoretical
understanding are improved.

Though we have shown, under rather standard assumptions, that the
$S$-matrix has an infinite number of energy poles for any exotic 
quantum numbers, this is only a necessary condition for exotic 
physical states to emerge. A sufficient condition would be to prove 
that some of those energy poles can be placed near the physical region. 
For such goal, however, we need to know more detailed dynamics.

\subsection{Phenomenological questions}

Similar problems, though apparently different, arise also in
phenomenological approaches. Consider, \textit{e.g.}, the case when
one extracts an amplitude from experimental data, and then
continues it into the complex energy plane to search for poles.
Evidently, experimental data are given at discrete energies (we
neglect experimental uncertainties for the moment). It is well
known, however, that analytical continuation from a discrete set
of points is ambiguous.  Therefore, the position of a pole found
from experimental data, and even its existence, formally speaking,
may be ambiguous as well. Quantitatively, such ambiguity is the smaller,
the nearer to experimental points is the complex energy value,
reached in the continuation. Thus, the problem of searching for
poles becomes most ambiguous for poles far from the physical
region. We see again that only states with not
very large widths are physically reliable.

Near the physical region, ambiguity of the continued amplitude should
be quantitatively small. Nevertheless, a pole may still be ambiguous,
if it has a sufficiently small residue, which corresponds to a small
total and/or partial decay width.

Up to now, we have considered the possible pole uncertainties as a
mathematical problem. However, experimental errors and ``technical''
methods used in extracting amplitudes may result in additional
ambiguities for the states of either very large or very small
width. For instance, as we discussed in Ref.~\cite{aasw3}, the
standard procedures for PWA may miss narrow resonances with, say,
$\Gamma^{tot} < 20-30$~MeV. Note, that the conventional PWA for elastic
scattering may miss also a resonance with $\Gamma^{tot} > 30$~MeV, if
it has small elastic partial width $\Gamma^{el}$, providing a small
elastic branching ratio of, say, less than 5\%.

Now, given an understanding of the possibility, and even necessity,
of complex-energy poles with various flavor quantum numbers, including
exotic ones, we are ready to discuss the present status of such poles
on the basis of existing experimental data.

\section{Resonance states in Partial-Wave Analyses}

As explained in the Introduction, we will search for $S$-matrix
poles with $S=+1,\,I=1$ on one side, and with $S=0,\,I=3/2$ on the
other.  Formally, this purpose could be achieved by studying only
scattering of $K^+p$ and $\pi^+p$ respectively. However, wider
sets of experimental data may be involved by studying all
available charge combinations of $KN$ and $\pi N$ scattering, with
later separation of isospin states. In this way one can use the
latest published PWA's for $KN$~\cite{KN} and $\pi N$~\cite{piN}
amplitudes correspondingly. We will discuss their application
separately for wide and narrow states.

\subsection{Wide states}

Energy-dependent PWA allows one to analytically continue the
partial-wave amplitudes into the complex-energy plane and to
search there for poles of the amplitudes. For $KN$ scattering with
isospin $I=1$, the latest PWA~\cite{KN}, up to $W=2650$~MeV, reveals
two poles having not too large imaginary parts:
$(M,\,\Gamma^{tot}/2)=(1811,\, 118)$~MeV for the $P_{13}$-wave,
and $(M,\,\Gamma^{tot}/2)=(2074,\,253)$~MeV for the $D_{15}$-wave.
We will consider these as candidates for the states $\Theta_1$ 
(the subscript denotes the isospin value $I=1$) with $J^P=3/2^+$ 
and $J^P=5/2^-$ correspondingly.

As explained in the Introduction, the $\Theta_1$, being a member
of \textbf{27}-plet, is to be accompanied by a $\Delta$-like 
partner.  For the two above states, such companions should have 
the same values of $J^P$.  Therefore, they are expected to be seen 
as poles in the $\pi N$ partial-wave amplitudes $P_{33}$ and $D_{35}$.

Indeed, the latest PWA for the $\pi N$ scattering~\cite{piN},
up to $W=2260$~MeV, contains the pole in the $D_{35}$ amplitude with
$(M,\,\Gamma^{tot}/2)=(1966,\,182)$~MeV. Quite reasonably, it may
be the $\Delta$-like companion for the $\Theta_1$ of
$(M,\,\Gamma^{tot}/2)=(2074,\,253)$~MeV with $J^P=5/2^-$.

What can be said about the $P_{33}$ amplitude?  The Review of
Particle Properties contains two candidates, the $P_{33}(1600)$
and $P_{33}(1920)$ with widths in the 200$-$400~MeV range~\cite{RPP}.
These are based on earlier PWA's, including previous VPI
results~\cite{Arnd} (where only $P_{33}(1600)$ was seen).
Note that evidence for $P_{33}$ states above the $P_{33}(1232)$ is
weak in elastic $\pi N$ scattering, possibly because of small,
$<20\%$, elastic branching ratios. The latest published
analysis of $\pi N$ data~\cite{piN} does not give reliable candidates
in the appropriate energy range.

One could assign either of the above states to be the $\Delta$-like
companion for the $\Theta_1$ of $(M,\,\Gamma^{tot}/2) =
(1811,\,118)$~MeV with $J^P=3/2^+$ (though both candidates have
masses largely separated from the $P_{13}$ state found in $KN$
scattering). However, we consider $P_{33}(1920)$ to be less reliable
(smaller elastic branching ratio, the mass is nearer to the upper
end of the energy ranges used in PWA's). Position of the pole for
$\Delta(1600)$, and even its existence, looks rather uncertain as 
well, possibly due to influence of nearby thresholds, $\pi\Delta$ 
and $\rho N$.  This state and its properties need confirmation that 
might come from detailed studies of inelastic processes (\textit{e.g.},
$\pi N$ and/or $\gamma N$ production of the $\pi\pi N$ final state).
At the present moment, we will tentatively use $\Delta(1600)$ with
PDG-values of its parameters~\cite{RPP}.

Summarizing the situation for wide resonances, we have two pairs
of the candidates for \textbf{27}-plet members:
\be
\label{32}
J^P=\frac{3}{2}^+,~~~(M_{\Theta_1},\,
\Gamma^{tot}_{\Theta_1})=(1811,\,236)~{\rm MeV},~~~ (M_{\Delta},
\,\Gamma^{tot}_{\Delta})=(1600,\,300)~{\rm MeV}; \ee
and
\be
\label{52}
J^P=\frac{5}{2}^-,~~~(M_{\Theta_1},\,\Gamma^{tot}_{\Theta_1})=
(2074,\,506)~{\rm MeV},~~~ (M_{\Delta},\,\Gamma^{tot}_{\Delta})
=(1966,\,364)~{\rm MeV}.
\ee

Note that the number of wide $\Delta$-states in the $\pi N$
analysis~\cite{piN} is larger than the number of wide
$\Theta_1$-states in the $KN$ analysis~\cite{KN}. While the pairs
$(\Theta_1,\Delta)$ are candidates for \textbf{27}-plets, the
excess $\Delta$'s may correspond to more familiar decuplets,
without any $\Theta_1$-companions.

We can recall here that every \textbf{27}-plet should contain also
one $N$-like state. Baryon tables~\cite{RPP}, indeed, demonstrate
possible wide candidates, even two ones for each of pairs (\ref{32})
and (\ref{52}).

For the pair (\ref{32}), with $J^P=3/2^+$, one candidate is the
$P_{13}$ state $N(1720)$ (with 4-star status~\cite{RPP}). It is
present also in the new solution FA02~\cite{piN}, with the pole
parameters $(M_{\rm pole},\, \Gamma^{tot}_{\rm pole}/2) =
(1655,\,139)$~MeV or the Breit-Wigner parameters $(M_{\rm BW},\,
\Gamma^{tot}_{\rm BW}/2)=(1750,\,128)$~MeV. The properties of
$N(1720)$ seem to agree with $\chi$SM-calculations~\cite{elkp,wm}
for the \textbf{27}-plet with $J^P=3/2^+$. Another candidate could
be the $P_{13}$ state $N(1900)$, if it exists (it has $\Gamma^{tot}
\approx500$~MeV and only 2 stars~\cite{RPP}; the solution
FA02~\cite{piN} does not contain this state).

For the pair (\ref{52}), with $J^P=\frac{5}{2}^-$, one candidate is
the $D_{15}$ state $N(1675)$. It also has 4-star PDG-status~\cite{RPP}
and is also present in the solution FA02, with the pole parameters
$(M_{\rm pole},\, \Gamma^{tot}_{\rm pole}/2)=(1659,\, 73)$~MeV and
the Breit-Wigner parameters $(M_{\rm BW},\, \Gamma^{tot}_{\rm
BW}/2)=(1676,\,76)$~MeV~\cite{piN}. The second candidate could
be one more $D_{15}$ state $N(2200)$, with $\Gamma^{tot}$ of
$300-400$~MeV and 2 stars~\cite{RPP}. It does not appear in the
FA02~\cite{piN} (probably as this is near the upper boundary of the
analysis).

\subsection{Narrow states}

As has been explained above (more detailed discussions see in
Refs.~\cite{aasw3,aasw4}), any PWA by itself tends to miss narrow
resonances. That is why we suggested~\cite{aasw3} to modify the PWA
procedure by explicitly assuming the existence of a narrow resonance
and comparing the quality of fits with and without this addition
(more details of the method and related formulas are in
Ref.~\cite{aasw4}). Such an approach was used initially to place
restrictions on light resonances in pion-nucleon
scattering~\cite{aasw3}.

This method was applied also to studies of $\Theta^+(1540)$~\cite{asw},
giving strict limitations on the $\Theta^+$ quantum numbers and,
especially, its width, confirmed by results of other approaches.
This method was then used~\cite{aasw4} to search $\pi N$ scattering
data for a narrow nucleon-like state assumed to be a member of the
antidecuplet, accompanying the $\Theta^+(1540)$. The two discovered
candidate mass regions have recently obtained preliminary experimental
support from direct measurements by the STAR and GRAAL
collaborations~\cite{kab,slav}. Thus, the modified PWA looks to be a
useful instrument in the search for narrow resonances.

Let us recall some features of this approach. To the canonical PWA
procedure, we add the explicit assumption that there is a resonance
with fixed parameters. We scan the assumed value of the resonance
mass, and also values of the total width and (for the inelastic
resonance) the elastic partial width (instead of the elastic partial
width, one may scan the elastic branching ratio). Then the database
is fitted with and without this resonance hypothesis. Evidently,
our procedure introduces additional parameters and, at first sight,
should always decrease $\chi^2$. However, this is not necessary true,
since those additional parameters are given with a specific
functional form (narrow resonance) that may be admissible for some
values of parameters, but definitely not for others.  We consider the
resonance to be possible if $\chi^2$ with the additional resonance is
smaller than without it.

Our experience shows that this approach becomes inefficient for
resonances with large total width $\Gamma^{tot}$ (if it is larger
than, say, 30~MeV). In this sense, our method appears to be
complementary with the conventional PWA, which is sensitive just
to resonances with larger values of $\Gamma^{tot}$. Moreover, for
an inelastic resonance, our approach is mainly sensitive to the upper
boundary for the elastic partial width $\Gamma^{el}$, but is not so
sensitive to the particular value of $\Gamma^{tot}$ (see discussion
in Ref.~\cite{aasw4}).

Here, we apply the modified PWA to search for possible members of a
\textbf{27}-plet, which could be seen as narrow resonances in
kaon-nucleon and pion-nucleon scattering. For the $KN$ scattering,
with the threshold near 1440~MeV, we take the c.m. energy interval
from 1500~MeV up to 1750~MeV, below the $K\Delta$ threshold. For the
$\pi N$ scattering, we take the same interval. We investigate $S$,
$P$, and $D$ partial-wave amplitudes, essentially involved in the
PWA's of Refs.~\cite{KN,piN}.  As a result, we find several
candidates in each of the waves, for both $KN$ and $\pi N$ scattering.
The elastic partial widths are restricted to very small values.  For
instance, the \mbox{$S_{31}$-wave} of $\pi N$ interaction suggests
three $\Delta$-like candidates having \mbox{$J^P=1/2^-$} and $(M,\,
\Gamma^{el})=(1570~{\rm MeV},\, <250~{\rm keV});\, (1630~{\rm
MeV},\,$\mbox{$<30~{\rm keV});$} $\, (1740~{\rm MeV},\,<90~{\rm keV}).$
There are candidates with even smaller $\Gamma^{el}$, down to 10~keV,
but we do not consider them here. In this respect, we recall once
more that our approach, generally, cannot prove existence of any narrow
resonance. Our modification of an amplitude has a definite functional
form (corresponding to a narrow resonance).  However, it may appear
successful even without any true resonance, \textit{e.g.}, as an
imitation of some other features of the amplitude, taken into account
in the conventional (unmodified) PWA, but with insufficient accuracy
(if considered at all, see discussion in Ref.~\cite{aasw3}). Therefore,
we consider our candidates only as evidence that the corresponding
values of energies (masses) are worth more detailed experimental
investigation.

After removing the most doubtful candidates, we have the set of
possible narrow resonances summarized in Table~\ref{tbl1} for the
$S$-wave ($J^P=1/2^-$), two $P$-waves ($J^P=1/2^+$ and $3/2^+$), and
two $D$-waves ($J^P=3/2^-$ and $5/2^-$). It is interesting that we
see equal numbers of candidates for both $\Delta$- and
$\Theta_1$-like narrow states, contrary to the case of wide
resonances.

\section{Discussion of the results}

Let us first summarize the results of the preceding Section. We do
see several correlated pairs of possible resonances $\Theta_1$ and
$\Delta$, each having the same spin-parity and nearby masses, which
may be considered in a natural way as members of \textbf{27}-plets.
Note that such an interpretation could be spoiled if we saw more
$\Theta_1$ than $\Delta$ candidates.  However, we have not
encountered this problem, though both the data sets and the analyses
for $KN$ and $\pi N$ scattering are totally independent. Indeed, an
excess of $\Delta$ candidates does not prevent the \textbf{27}-plet
hypothesis, since the excess $\Delta$'s may be related to other
flavor multiplets, \textit{e.g.}, to decuplets having no open exotics.
For excess $\Theta_1$'s, no reasonable multiplet prescription would
be possible.

Further, we see two classes of candidates. The first contains very
broad states, with total widths of hundreds MeV. The second is
restricted to very narrow widths.  Here the elastic width $\Gamma^{el}$
could be as small as 100~keV, or even tens of keV. These values
coincide with the total widths $\Gamma^{tot}$ of $\Theta_1$-like
candidates having masses in the elastic region (below $\sim1570$~MeV).
Other $\Theta_1$-like candidates, with higher masses, and all
$\Delta$-like candidates, may have $\Gamma^{tot}>\Gamma^{el}$.
However, we study the $\Theta_1$-masses below the $\Delta K$ and
$NK^\ast$ thresholds, and expect the inelastic contribution to be
moderate. Thus, though we cannot reliably extract the total widths
of these narrow candidates, our procedure, in any case, suggests
that they, most probably, should have $\Gamma^{tot}$ small, not
larger than low tens of MeV.

Let us compare our candidates with theoretical expectations.  The
$\Theta^+$ and the whole antidecuplet with $J^P=1/2^+$ were
predicted~\cite{dpp} on the basis of the chiral soliton model
($\chi$SM). Definitely, such an approach also predicts two
\textbf{27}-plets, one with $J^P=3/2^+$ and one with $J^P=1/2^+$
(see \textit{e.g.}, Refs.~\cite{wk,elkp}; note that the case
of $J^P=1/2^+$ has usually been discussed much more briefly).
Instead, among our candidates there are even several correlated
pairs ($\Theta_1,\Delta$) with $J^P=3/2^+$, several pairs with
$J^P=1/2^+$, and also pairs with different combinations of spins
and parities, not considered in $\chi$SM. All those pairs are
expected to label the corresponding \textbf{27}-plets.

Properties of our candidates for \textbf{27}-plet members look
different from the conventional $\chi$SM expectations. For
instance, estimations in $\chi$SM~\cite{elkp} for the \textbf{27}-plet
with $J^P=3/2^+$ have given the $\Theta_1$ to be about 60~MeV
heavier than the antidecuplet $\Theta^+$, and its $\Delta$
companion about 50~MeV heavier than the $\Theta_1$. Similar
predictions were also given in other papers, \textit{e.g.}, in
Refs.~\cite{wk,wm}.

Instead, the $\Theta_1$-state in our wide candidate pair of
Eq.~(\ref{32}) is heavier than the $\Theta^+$ by more than
250~MeV. The corresponding $\Delta$-state could be lighter or heavier
than the $\Theta_1$-state by about a hundred MeV, depending on the
assignment. For narrow pairs (in particular, for the pairs with
$J^P=3/2^+$), as shown in \mbox{Table~\ref{tbl1}}, masses of the
$\Theta_1$ and $\Delta$ companions are nearly the same.

Thus, none of our candidate pairs correspond to the expected
($\Theta_1, \Delta)$-mass ordering. Evidently, the $\Theta_1$-masses,
with respect to the $\Theta^+$, are also different from expectations.
Note, that such conclusions would not change if we used other pair
combinations of $\Theta_1$'s and $\Delta$'s having the same
$J^P$-values.

Expectations for widths of the \textbf{27}-plet baryons have also been
discussed in the literature, though mainly for explicitly exotic
states. The width of the $\Theta_1$ with $J^P=3/2^+$ was estimated
in $\chi$SM to be between 37 and 66~MeV~\cite{elkp,wm}. Our candidates,
again, do not demonstrate the expected values. They have widths either
essentially wider (more than 200~MeV), or essentially narrower (less
than 1~MeV). Width estimations for the $\Delta$-member of the same
\textbf{27}-plet, $\Gamma^{tot}\sim100$~MeV and some tens MeV for the
partial width $\Gamma^{el}$ of the $\pi N$ decay~\cite{wm}, also do
not correspond to the properties of our candidates.

The current literature suggests some other, alternative to $\chi$SM,
approaches to describe members of higher baryon multiplets. For
example, the QCD sum rules~\cite{sr} predict $\Theta_1$ (and even
$\Theta_2$, with $I=2$, belonging to \textbf{35}) to have
$J^P=1/2^-,\,3/2^-$ and masses nearly the same as the $\Theta^+$,
though positive parities are not excluded. It appears, however, that
the suggested alternative approaches similarly fail to describe our
candidates.

Nevertheless, we do not insist that the $\chi$SM (or any other model)
is incorrect for higher flavor multiplets. The problem is that the
published calculations in the framework of a particular model always
use some additional assumptions, not inherent in the model itself.
For the $\chi$SM, states considered in the literature are rotational
excitations of the soliton.  As a result, all baryon states in the
$\chi$SM publications have only positive parities, though negative
parity baryon resonances are certainly known to exist in
nature~\cite{RPP}.
Excitations of other types, vibrational for example, are also possible,
though methods for their study have not been elaborated. Note, that
the vibrational excitations might provide a more numerous set of
states than the rotational ones. Such states could have opposite
parities and essentially smaller widths as compared to the states
with rotational excitations.

In any case, existence of two kinds of states, very wide and very
narrow, would hint at a very interesting dynamics. Note that
similar effects emerge even for the rotational excitations in the
$\chi$SM.  Indeed, the coupling of decuplet baryons to octet states
is well described and provides rather large width of, say,
$\Delta(1232)$.  On the other hand, theoretical estimates of the
antidecuplet-octet coupling reveal some suppression~\cite{dpp},
though its origin has not yet been understood (experimentally, the
suppression seems to be even stronger than expected). A similar
situation might arise also for the higher flavor multiplets.
An interesting fact in this respect is that among our narrow
candidates there are no excess $\Delta$-like states, which could
belong to decuplets. This might mean that only higher multi-quark
multiplets can have suppressed hadronic widths.

In discussing masses and widths of the candidate states, one should
take into account the influence of their mixing(s) with states having
the same flavor and spin-parity quantum numbers, but belonging to
other flavor multiplets. Such mixing may essentially shift the
expected masses and decay properties of multi-quark baryons, as has
been suggested and demonstrated, \textit{e.g.}, in
Refs.~\cite{aasw4,jw,dp,gp}. The large number of vibrationally
excited states might be very important in this respect.

We should again emphasize that our results give evidence for candidate
states, but cannot prove their existence. Therefore, direct
experimental searches, with good precision and good mass resolution,
become important in the final decision. As for the $\Theta^{++}$,
several searches for this state in $K^+p$ mass distributions have
been published~\cite{henry,elsa,clas2,herm,zeus,bab}, all with
negative results. This is quite natural, having in mind the widths of
our candidates. Indeed, it would be difficult to separate from
background a bump with the width of hundreds of MeV, which should
correspond to our wide candidates for $\Theta_1$ in pairs (\ref{32})
and (\ref{52}). On the other hand, manifestation of a peak with a
very small width, $<1$~MeV, especially with small production cross
section, would be essentially suppressed by the experimental
resolution $\sim10$~MeV.

To conclude, we have argued that the $S$-matrix (and, thus, the
corresponding amplitudes) should have poles with any quantum numbers,
exotic or non-exotic. Thus, a necessary condition for physical exotic
states to exist is satisfied, though we can not prove that the poles
reveal themselves indeed as exotic hadrons. Nevertheless, our result
presents a new argument for such a possibility. When studying the $KN$
and $\pi N$ scatterings, we see in their PWA's the baryon candidates,
$\Theta_1$'s and $\Delta$'s, for members of several multi-quark
\textbf{27}-plets. Properties of the candidates (their quantum numbers,
masses and widths) do not quite corespond to expectations published 
in the literature. This might have natural explanation, but existence 
of our candidates should still be checked more reliably by direct 
measurements, with good accuracy and resolution.

\section*{Acknowledgments}
We thank V.~Yu.~Petrov, M.~V.~Polyakov, and M.~Praszalowicz for
valuable discussions.  The work was partly supported by the
U.~S.~Department of Energy Grant DE--FG02--99ER41110, by the
Jefferson Laboratory, by the Southeastern Universities Research
Association under DOE Contract DE--AC05--84ER40150, by the
Russian-German Collaboration (DFG, RFFI), by the
COSY-Juelich-project, and by the Russian State grant
RSGSS-1124.2003.2.


\begin{table}[th]
\caption{Candidates for narrow $\Delta$- and $\Theta_1$-like
         states in $\pi N$ and $KN$ scattering. \label{tbl1}}
\vspace{0.15in}
\begin{tabular}{l|c|c}
\colrule
J$^P$  & (M$_\Delta$, $\Gamma_\Delta^{el}$) [MeV, keV]
       & (M$_{\Theta 1}$, $\Gamma_{\Theta 1}^{el}$) [MeV, keV] \\
\colrule
1/2$^-$& (1570, $<$250) & (1550,  $<$80) \\
       & (1630,  $<$30) & (1640, $<$100) \\
       & (1740,  $<$90) & (1740,  $<$60) \\
\colrule
1/2$^+$& (1550, $<$400) & (1530, $<$100) \\
       & (1680,  $<$50) & (1660,  $<$80) \\
       & (1730,  $<$30) & (1740, $<$100) \\
\colrule
3/2$^+$& (1550, $<$100) & (1530,  $<$80) \\
       & (1660,  $<$30) & (1650,  $<$50) \\
       & (1720,  $<$70) & (1710,  $<$30) \\
\colrule
3/2$^-$& (1520,  $<$50) & (1530, $<$150) \\
       & (1570, $<$120) & (1570,  $<$70) \\
       & (1730,  $<$60) & (1740,  $<$80) \\
\colrule
5/2$^-$& (1510,  $<$50) & (1530,  $<$70) \\
       & (1570,  $<$30) & (1570,  $<$60) \\
       & (1620,  $<$15) & (1640, $<$100) \\
       & (1700,  $<$70) & (1680,  $<$60) \\
\colrule
\end{tabular}
\end{table}
\vspace{4in}
\end{document}